\documentclass[showpacs,preprintnumbers,amsmath,amssymb, pr, twocolumn]{revtex4}

\usepackage{stmaryrd}
\usepackage{txfonts}
\usepackage{amssymb}
\usepackage{mathrsfs}
\usepackage{graphicx}
\usepackage{dcolumn}
\usepackage{bm}
\usepackage{epsfig}
\begin{document}

\title{Vortex States and Phase Diagram of Multi-component Ginzburg-Landau Theory with
Competing Repulsive and Attractive Vortex Interactions}

\author{Shi-Zeng Lin and Xiao Hu}

\affiliation{WPI Center for Materials Nanoarchitectonics, National
Institute for Materials Science, Tsukuba 305-0044, Japan
}

\begin{abstract}
We investigate the behavior of vortices of multi-component
superconductivity, realized in $\rm{MgB_2}$ and Fe-based
superconductors, within the framework of Ginzburg-Landau (GL) theory in
terms of numerical calculations of the time-dependent
GL equations and the variational method. It is revealed
that close to the critical point of the composite system the
inter-component coupling makes the system behave as a single
component superconductivity in most cases. However, when the bare
mean-field critical points of the two components coincide with each
other, and furthermore the inter-band coupling disappears at the
same temperature, interesting phenomena occur as follows. Vortices
interact attractively at large separation and repulsively at short
distance in certain parameter space. Because of the non-monotonic
interaction profile, phase separations between vortex clusters of
triangular order and the Meissner state take place, which
indicates a first-order phase transition associated with the
penetration of the magnetic field into a superconductor
sample. Phase diagrams of vortex states are then
constructed with the associated magnetization curve. It is found
that all these behavior interpolates the features of the type I and
II superconductors.
\end{abstract}


\date{\today}

 \maketitle
\section{Introduction}

A quantized vortex is a topological excitation of superfluidity and
superconductivity as the hallmark of a single wave function for the
macroscopic quantum state. Revealed first by Abrikosov based on the
celebrated Ginzburg-Landau (GL) theory, a vortex in superconductor
carries a normal core with size of the order of
superconducting coherence length $\xi$, and a quantized magnetic
flux distributed over the London penetration depth $\lambda$; in
superconductors categorized as type II with
$\kappa=\lambda/\xi>1/\sqrt{2}$, as opposed to type I with
$\kappa<1/\sqrt{2}$, vortices repel each other because of the
circular supercurrents and thus form a triangular vortex lattice. In
type I superconductors, vortices would attract each other in order
to gain the superconducting condensation energy, and thus collapse
into a continuum of normal region at equilibrium.

 The recently discovered
multi-band superconductors, such as $\rm{MgB_2}$ \cite{Nagamatsu01}
and iron pnictide superconductors\cite{Kamihara08}, can exhibit
novel
phenomena\cite{Tanaka02,Babaev02,Babaev07,Babaev09,Chibotaru10,Geurts10,Blumberg07}
(see Ref.\cite{Xi08} for a review), without counterpart in
single-band superconductors. It was first proposed theoretically by
Babaev and Speight\cite{Babaev05,Babaev10} that, when the
London penetration depth falls in between the two coherence
lengths, vortices may attract at large separation and repel at
short distance because of the competition of different length scales
in different condensates. It was then discussed by Moshchalkov
\emph{et al.} that this situation was realized in their sample of
$\rm{MgB_2}$, as manifested by the unconventional stripe and
gossamer-like vortex patterns\cite{Moshchalkov09}. They coined the
term of type 1.5 superconductor for the class of multi-band
superconductors.

While this scenario is intriguing and has attracted considerable
interests \cite{Geyer10,Dao11,Brandt10}, there are several points
waiting for further investigation. In the theoretical side, one notices
that the original proposal was based on weak inter-component
coupling limit, and the discussions were implicitly extended to
low-temperature regime. In doing so, one should always keep in mind
that GL theory works only close to the critical point of the
composite system, at least in principle. Any result appearing only
away from the critical point needs a careful and independent check.
In the experimental side, the observed inter-vortex separation $\sim
2\mu$m is much larger than the distance associated with the possible
energy minimum estimated by the theory \cite{Babaev05}, lacking a
consistent understanding. The random vortex configurations reported
in Refs.~\cite{Moshchalkov09, Nishio10PRB} seem incompatible with
common experiences in material science: particles governed by the
Lennard-Jones interactions form solids with regular orders, and thus
those random patterns are hard to be considered as a
property of equilibrium at usual conditions in absence of random pins.

In the present work, we reveal first a case that the bare
mean-field critical points of the two components coincide with each
other, and furthermore the inter-band coupling disappears at the
same temperature. We introduce two vortices into a square
superconductor by setting appropriate boundary condition, and find
with the approach of time-dependent GL (TDGL) that, for appropriate
parameters in the GL free energy functional, the two vortices are
separated by a distance associated with an energy minimum, which
does not change with the system size for simulations. This verifies the attraction
between vortices at large distance and repulsion at short distance.
The vortices are found stable against thermal fluctuations. We
evaluate the full dependence of interaction potential on vortex
separation by the variational method. Distribution of
vortices are then simulated based on the Langevin dynamics. Instead of
uniform vortex lattice in type II superconductors as well as the
lamella structure in type I superconductors, we observe phase
separations among clusters of triangularly ordered vortices and
Meissner regions, which indicates a first-order phase transition
associated with penetration of vortices into the system upon
increase of external magnetic field. Based on these observations we
construct the phase diagram for superconductors with the novel
vortex interaction.

We also investigate the vortex interaction when two
components have different bare critical points and a finite
inter-band coupling. Close to the critical point of the composite
system, the superconductivity in different components is strongly
correlated and only one divergent length associated with the
variation of order parameters at the vortex core exists. Thus
vortices are purely repulsive or attractive close to the
critical point.

It is noticed that novel vortex attractions were discussed 40 years
ago in single-band superconductors, which caused phase separation
between Meissner phase and vortex lattice, and discontinuous jumps
in magnetization, especially at low temperature (see
Ref.\cite{Brandt95} for review). These superconductors are
characterized by small GL numbers, and the attractive vortex
interactions were attributed to correction to the GL approach from
the BCS theory\cite{Eilenber69,Jacobs71}, which is different from
the system discussed by \cite{Babaev05,Babaev10} and addressed in
the present work.

The remaining part of the paper is organized as follows. In Sec. II,
we present the free energy functional and the structure of a
vortex. In Sec. III, we present the results obtained by numerical
simulations of TDGL. In Sec. IV, the interaction potential is
obtained by the variational method. In Sec. V, the vortex
configuration is simulated based on the Langevin dynamics.
Sec. VI is devoted discussions, and the paper is concluded by
Sec. VII.

\section{Model}
The GL free energy functional with interband scattering is given by
\begin{equation}\label{eqs1n1}
\begin{array}{l}
\mathcal{F}=\sum _{i=1,2}\left[\alpha _i\left|\Psi _i\right|{}^2+\frac{\beta _i}{2}\left|\Psi _i\right|{}^4+\frac{1}{2m_i}\left|\left(-i \hbar \nabla  -\frac{2e}{c}\mathbf{A}\right)\Psi _i\right|{}^2\right]\\
+\frac{1}{8\pi }(\nabla \times \mathbf{A})^2-\gamma  \left(\Psi _1^*\Psi _2+\Psi _2^*\Psi _1)\right.,
\end{array}
\end{equation}
\noindent with all symbols conventionally defined\cite{TinkhamBook}.
Here we focus on the Josephson-like inter-band coupling with
strength $\gamma$, noticing that the main results remain
qualitatively the same even including other possible
interactions. For $\rm{MgB_2}$ $\gamma>0$, while for iron-based
superconductors, $\gamma$ is presumably negative\cite{Mazin08}. The
temperature dependence of the quadratic terms are
$\alpha_i(T)=\alpha_i(0)(1-T/T_c)$ with $\alpha_i(0)<0$, and
$\gamma(T)=\gamma(0)(1-T/T_c)$. We notice that in this case the
physics is the same for the whole temperature regime starting from
$T_c$ except a renormalization of length. We leave the discussion on
more general cases to Sec. VI.

For convenience, we define the following lengths
$\lambda_i=\sqrt{{m_ic^2\beta_i}/{8\pi |\alpha_i| e^2}}$ and
$\xi_i=\sqrt{{\hbar ^2}/{2m_i|\alpha_i|}}$, the penetration depth
and coherence lengths in the respective single-band condensates
($\gamma=0$). For the present interest, we adopt the coefficients in
the GL free energy functional Eq. (\ref{eqs1n1}) such that at $T=0$,
$\xi_1=51$nm, $\lambda_1=25$nm, $\xi_2=8$nm, and $\lambda_2=30$nm
at very low temperatures.
The inter-band coupling is $\gamma(0)=-0.4\alpha_1(0)$.

We introduce the dimensionless quantities for convenience,
\begin{equation}\label{eqs1n2}
\begin{array}{l}
x=\lambda _1x',\ \ \Psi _i=\Psi _{10}\Psi _i', \ \ \mathbf{A}=\lambda _1H_{1c}\sqrt{2}\mathbf{A}',\ \ \mathcal{F}=\frac{H_{1c}^2}{4\pi} \mathcal{F}',\\
\gamma =\gamma '\left|\alpha _1\right|,\ \  \mathbf{B}=H_{1c}\sqrt{2}\mathbf{B}',\ \  \mathbf{J}=\frac{2e\hbar \Psi _{10}^2}{m_1\xi _1}\mathbf{J}',
\end{array}
\end{equation}
where $\Psi _{10}^2=\left|\alpha _1\right|\left/\beta _1\right.$ is the bulk value, $H_{1c}=\sqrt{4\pi \left|a_1\right|\Psi _{10}^2}$ the thermodynamic critical field of the first condensate. $\mathbf{B}$ is the magnetic induction and $\mathbf{J}$ is the supercurrent.

For clarity, we drop the prime in the dimensionless quantities in the following calculations. Then we have the free energy in the dimensionless units
\begin{equation}\label{eqs1n3}
\begin{array}{l}
\mathcal{F}=\sum _{i=1,2}\left[\frac{\alpha _i}{\left|\alpha _1\right|}\left|\Psi _i\right|{}^2+\frac{\beta _i}{2\beta _1}\left|\Psi _i\right|{}^4+\frac{m_1}{m_i}\left|\left(\frac{1}{i \kappa_1 }\nabla -\textbf{A}\right)\Psi _i\right|^2\right]\\
+(\nabla \times \textbf{A})^2 -\gamma (\Psi _1^*\Psi _2+\Psi _2^*\Psi _1),
 \end{array}
\end{equation}
where $\kappa _1=\lambda _1/\xi _1$ is the GL parameter.

Minimizing $\mathcal{F}$ with respect to $\Psi _i$ and $\mathbf{A}$, we obtain the GL equations
\begin{equation}\label{eqs1n4}
-\Psi _1+\left|\Psi _1\right|{}^2\Psi _1+\left(\frac{1}{i \kappa _1}\nabla  -\textbf{A}\right){}^2\Psi _1-\gamma  \Psi _2=0,
\end{equation}
\begin{equation}\label{eqs1n5}
-\frac{\alpha _2}{\alpha _1}\Psi _2+\frac{\beta _2}{\beta _1}\left|\Psi _2\right|{}^2\Psi _2+\frac{m_1}{m_2}\left(\frac{1}{i \kappa _1}\nabla  -\textbf{A}\right){}^2\Psi _2-\gamma  \Psi _1=0,
\end{equation}
\begin{equation}\label{eqs1n6}
\begin{array}{l}
\nabla \times \nabla \times \textbf{A}=\frac{1}{2i \kappa _1}(\Psi _1^*\nabla \Psi _1-\Psi _1\nabla \Psi _1^*)-\left|\Psi _1\right|{}^2\textbf{A}
\\+\frac{m_1}{m_2}\left(\frac{1}{2i \kappa _1}(\Psi _2^*\nabla \Psi _2-\Psi _2\nabla \Psi _2^*)-\left|\Psi _2\right|{}^2\textbf{A}\right).
\end{array}
\end{equation}
Equation (\ref{eqs1n6}) describes the screening of the magnetic field by the superconducting condensates. Using the London approximation, the effective London penetration depth for two-band superconductors is
\begin{equation}\label{eqs1n7}
\lambda =1\left/\sqrt{\left|\Psi _{10}\right|{}^2+\frac{m_1}{m_2}\left|\Psi _{20}\right|{}^2}\right.,
\end{equation}
where $\Psi_{i0}$ is the bulk value of the $i$th superconducting condensate. The response of two-band superconductors to magnetic fields is described by a single length scale $\lambda$, because both condensates couple to the same gauge field. For $\gamma=0$, we have $\Psi _{10}=1$ and $\Psi _{20}=\sqrt{{\alpha _2\beta _1}/{\alpha _1\beta _2}}$. The interband coupling shifts the bulk value and thus modifies the corresponding penetration depth.

Next we construct a vortex solution to Eqs. (\ref{eqs1n4}),
(\ref{eqs1n5}) and (\ref{eqs1n6}). It is observed that the
two condensates should have the same vorticity and phase around the
vortex core to save energy if $\gamma>0$, while for $\gamma<0$ the
phase shift should be $\pi$. For condensates with different winding
number, the vortex is fractional quantized and the energy diverges
logarithmically with vortex size in bulk\cite{Babaev02}, and thus
thermodynamically unstable. Without loss of generality, we focus on
the positive $\gamma>0$ as in the case of
$\rm{MgB_2}$\cite{Nagamatsu01}. Presuming a straight vortex
line, and we look for a vortex with the following structure
\begin{equation}\label{eqs1n8}
\Psi _i=f_i(r)e^{i n \theta } \text{\    and\    } \textbf{A}=\frac{n a(r)}{\kappa _1 r}\textbf{e}_{\theta },
\end{equation}
where $r$ is the distance from the vortex core, $\textbf{e}_{\theta }$ the unit vector in the azimuthal direction and $n$ the vorticity. Substituting the ansatz into Eqs. (\ref{eqs1n4}), (\ref{eqs1n5}) and (\ref{eqs1n6}), we have
\begin{equation}\label{eqs1n9}
-f_1(r)+f_1^3(r)-\frac{1}{\kappa_1 ^2}\left(\partial _r^2f_1+\frac{1}{r}\partial _rf_1\right)+\frac{n^2( a-1)^2}{\kappa_1 ^2r^2}f_1-\gamma  f_2=0,
\end{equation}
\begin{equation}\label{eqs1n10}
\begin{array}{l}
-\frac{\alpha _2}{\alpha _1}f_2(r)+\frac{\beta _2}{\beta _1}f_2^3(r)\\
+\frac{m_1}{m_2}\left(-\frac{1}{\kappa_1 ^2}\left(\partial _r^2f_2+\frac{1}{r}\partial _rf_2\right)+\frac{n^2( a-1)^2}{\kappa_1 ^2r^2}f_2\right)-\gamma  f_1=0,
\end{array}
\end{equation}

\begin{equation}\label{eqs1n11}
\partial _r^2a-\frac{1}{r}\partial _ra+\left(f_1^2+\frac{m_1}{m_2}f_2^2\right)(1-a)=0.
\end{equation}
In the limit of $r\rightarrow\infty$, the wave functions recover the
bulk values $f_{10}$ and $f_{20}$. Defining $f_{20}=\eta f_{10}$
with $\eta>0$, we have the equations for $f_{10}$ and $\eta$

\begin{equation}\label{eqs1n12}
-1+f_{10}^2-\gamma  \eta  =0,
\end{equation}
\begin{equation}\label{eqs1n13}
-\frac{\alpha _2}{\alpha _1}\eta  +\frac{\beta _2}{\beta _1}\eta ^3 (1+\gamma  \eta  )-\gamma  =0.
\end{equation}

The radial variation of the wave functions and vector potential in
the asymptotic region of $r\rightarrow\infty$ can be found and are
given by
\begin{equation}\label{eqs1n14}
f_1=\sqrt{1+\gamma  \eta }+c_{\text{f1}}\exp \left(-\frac{r}{\sqrt{2}\xi_v}\right),
\end{equation}

\begin{equation}\label{eqs1n15}
f_2=\sqrt{\frac{\beta _1}{\beta _2}\left(\frac{\alpha _2}{\alpha _1}+\frac{\gamma }{\eta }\right)}+c_{\text{f2}}\exp
\left(-\frac{r}{\sqrt{2}\xi_v}\right),
\end{equation}
\begin{equation}\label{eqs1n16}
a=1+c_{\text{a}}\exp \left(-\frac{r}{\lambda_v}\right).
\end{equation}
At large distance, there is only one length scale for the two condensates. The penetration depth can be obtained straightforwardly
\begin{equation}\label{eqs1n17}
\lambda_v=1\left/\sqrt{\frac{m_1}{m_2}\frac{\beta _1}{\beta _2}\left(\frac{\alpha _2}{\alpha _1}+\frac{\gamma }{\eta }\right)+(1+\gamma  \eta )}\right..
\end{equation}
To calculate the coherence length, we substitute Eqs. (\ref{eqs1n14}) and (\ref{eqs1n15}) into Eqs. (\ref{eqs1n9}) and (\ref{eqs1n10}) and linearize the equations. $\xi_v$ is equivalent to the length scale of the small fluctuations in the bulk, and is given by the largest solution to the equation
\begin{equation}\label{eqs1n17a}
\left(2+3\gamma  \eta-\frac{1}{2\kappa_1 ^2\xi_v ^2}\right)\left(2 \frac{\alpha _2}{\alpha _1}+3\frac{\gamma} {\eta} -\frac{m_1}{m_2}\frac{1}{2\kappa_1 ^2\xi_v ^2}\right)-\gamma ^2=0.
\end{equation}
It is clear that the interband scattering modifies the coherence
length and penetration depth in a nontrivial way.

\section{TDGL approach}

To study the interaction between vortices, the structure of the
nonlinear vortex core is important and numerical calculations are
necessary. We first calculate the structure of vortex and their
interaction by minimizing the free energy in terms of TDGL equations
defined in the following way \cite{Xi08,Koshelev05}
\begin{equation}\label{eqs4n1a}
 \frac{{{\hbar ^2}}}{{2{m_i}D_i}}({\partial _t} + i\frac{{{2e}}}{\hbar }\Phi )\Psi_i   =   - \frac{{\delta \mathcal{F}}}{{\delta {\Psi_i ^*}}}+\zeta_i,
\end{equation}
\begin{equation}\label{eqs4n1b}
 \frac{\sigma }{c}(\frac{1}{c}{\partial _t}{\bf{A}} + \nabla \Phi )  =   - \frac{{\delta \mathcal{F}}}{{\delta {\bf{A}}}}+\mathbf{\zeta}_A,
\end{equation}
with
$D_i$ the diffusion constant, $\sigma$ the normal conductivity, and $\Phi$ the electric potential. By choosing a proper gauge, we can take $\Phi=0$.  $\zeta_j$ represents thermal noises satisfying $\left\langle\zeta_j\right\rangle=0$ and $\left\langle\zeta_j(\textbf{x}_1, t_1)\zeta_j(\textbf{x}_2,t_2)\right\rangle=\Gamma_j\delta(\textbf{x}_1-\textbf{x}_2)\delta(t_1-t_2)$, with $j=1,2 $ and $A$.
Since we are primarily interested in the equilibrium properties,
detailed dynamic relaxation process is irrelevant. Here, time is in units of ${\tau _1} = \xi _1^2/{D_1}$, the normal conductivity $\sigma$ in units of ${{{c^2}{\tau _1}}}/{{4\pi \lambda _1^2}}$.

\begin{figure}[t]
\psfig{figure=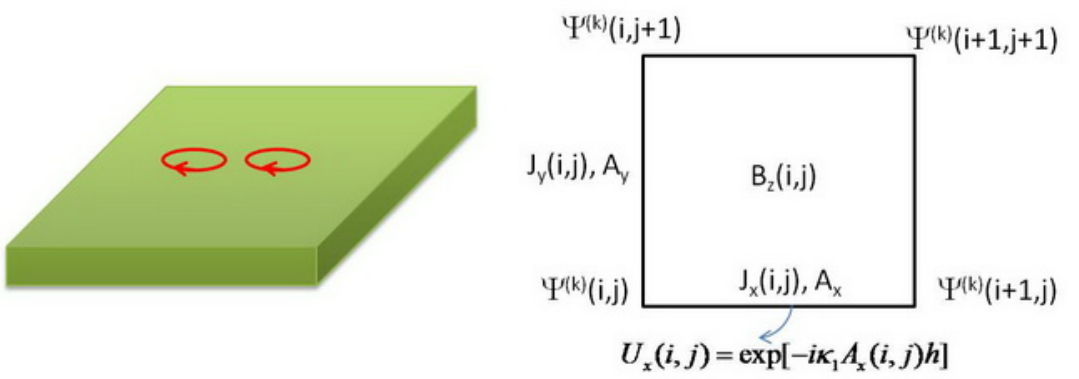,width=\columnwidth} \caption{\label{sf3}
(color online). Left: schematic view of two vortices in a
superconducting square disk. Right: numerical discretization scheme.
$\Psi^{(k)}$ is defined on nodes, $\mathbf{J}$, $\mathbf{A}$ and
$\mathbf{U}$ are defined on bonds, and $B_z$ is defined inside the
plaquette.}
\end{figure}

\subsection{Numerical techniques}

In order to integrate the TDGL equations, we generalize the
numerical method developed for single component
superconductors\cite{Gropp96} to two-component superconductors and
solve the TDGL equations Eqs. (\ref{eqs4n1a}) and (\ref{eqs4n1b})
numerically. For the parameters we are interested, the line tension
of vortices is high, therefore we approximate vortices as straight
lines. The problem is then simplified into two dimensions. To
maintain the gauge invariance after discretization\cite{Kaper95}, we
use the link variable defined as
 \begin{equation}\label{eqs4n1}
{U_\mu }(x,y) = \exp \left( - i\kappa_1 \int\limits_{{\mu _0}}^\mu  {{A_\mu }({\xi _\mu })d{\xi _\mu }} \right),
\end{equation}
where $\mu$ is either $x$ or $y$. Then the TDGL equations can be rewritten as
 \begin{equation}\label{eqs4n2}
{\partial _t}{\Psi _1} - \frac{1}{{{\kappa_1 ^2}}}\sum\limits_{\mu  = x,y} {U_\mu ^*\partial _\mu ^2({U_\mu }{\Psi _1})}  - {\Psi _1} + |{\Psi _1}{|^2}{\Psi _1} - \gamma {\Psi _2}-\zeta_1 = 0,
\end{equation}

 \begin{equation}\label{eqs4n3}
\begin{array}{l}
\frac{{{m_1}{D_1}}}{{{m_2}{D_2}}}{\partial _t}{\Psi _2} - \frac{{{m_1}}}{{{m_2}}}\frac{1}{{{\kappa_1 ^2}}}\sum\limits_{\mu  = x,y} {U_\mu ^*\partial _\mu ^2({U_\mu }{\Psi _2})}  \\
+ \frac{{{\alpha_2}}}{{|{\alpha_1}|}}{\Psi _2} + \frac{{{\beta_2}}}{{{\beta_1}}}|{\Psi _2}{|^2}{\Psi _2} - \gamma {\Psi _1} -\zeta_2 = 0,
\end{array}
\end{equation}

 \begin{equation}\label{eqs4n4}
\sigma {\partial _t}{\bf{A}} + \nabla  \times \nabla  \times {\bf{A}} = {{\bf{J}}}+\mathbf{\zeta}_A,
\end{equation}

 \begin{equation}\label{eqs4n5}
\begin{array}{l}
{J_{\mu }} = \frac{1}{{2i\kappa_1 }}[U_\mu ^*\Psi _1^*{\partial _\mu }({U_\mu }{\Psi _1}) - {U_\mu }{\Psi _1}{\partial _\mu }(U_\mu ^*\Psi _1^*)] \\
+ \frac{{{m_1}}}{{{m_2}}}\left( {\frac{1}{{2i\kappa_1 }}[U_\mu ^*\Psi _2^*{\partial _\mu }({U_\mu }{\Psi _2}) - {U_\mu }{\Psi _2}{\partial _\mu }(U_\mu ^*\Psi _2^*)]} \right).
\end{array}
\end{equation}
The disk is partitioned into square meshes of size $h$ as
schematically shown in Fig. \ref{sf3}. The magnetic field in the
plaquette with the index $(i, j)$ is
 \begin{equation}\label{eqs4n6}
{B_{z;\ i,j}} = \frac{{1 - {W_{z;\ i,j}}}}{{i\kappa_1 h^2}} \text{\ \ with\ \ }
{W_{z;\ i,j}} = U_{x;\ i,j + 1}^*U_{y;\ i,j}^*{U_{x;\ i,j}}{U_{y;\ i + 1,j}},
\end{equation}
and the supercurrent
 \begin{equation}\label{eqs4n7}
\begin{array}{l}
{J_{x;\ i,j}} = \frac{1}{{2i\kappa_1 {h}}}({U_{x;\ i,j}}\Psi _{i,j}^{(1)*}{\Psi _{i + 1,j}^{(1)}} - U_{x;\ i,j}^*{\Psi _{i,j}^{(1)}}\Psi _{i + 1,j}^{(1)*})\\
+\frac{m_1}{{2i\kappa_1 {h}m_2}}({U_{x;\ i,j}}\Psi_{i,j}^{(2)*}{\Psi_{i + 1,j}^{(2)}} - U_{x;\ i,j}^*{\Psi_{i,j}^{(2)}}\Psi _{i + 1,j}^{(2)*}),
\end{array}
\end{equation}
where $\Psi^{(k)}$ denotes the $k$th condensate. The $y$ component is obtained similarly. After the discretization, the TDGL equations become
\begin{widetext}
\begin{equation}\label{eqs4n8}
\begin{array}{l}
{\partial _t}\Psi _{i,j}^{(1)} = (1 - |\Psi _{i,j}^{(1)}{|^2})\Psi _{i,j}^{(1)} + \gamma \Psi _{i,j}^{(2)}+\zeta _{i,j}^{(1)}
+ \frac{1}{{{\kappa_1 ^2}}}\left( {\frac{{{U_{x;\ i,j}}\Psi _{i + 1,j}^{(1)} + U_{x;\ i - 1,j}^*\Psi _{i - 1,j}^{(1)} - 2\Psi _{i,j}^{(1)}}}{{h^2}}}+{\frac{{{U_{y;\ i,j}}\Psi _{i,j+1}^{(1)} + U_{y;\ i,j-1}^*\Psi _{i,j-1}^{(1)} - 2\Psi _{i,j}^{(1)}}}{{h^2}}}  \right),
\end{array}
\end{equation}

\begin{equation}\label{eqs4n9}
\begin{array}{l}
\frac{{{m_1}{D_1}}}{{{m_2}{D_2}}}{\partial _t}\Psi _{i,j}^{(2)} = ( - \frac{{{\alpha _2}}}{{|{\alpha _1}|}} - \frac{{{\beta_2}}}{{|{\beta_1}|}}|\Psi _{i,j}^{(2)}{|^2})\Psi _{i,j}^{(2)}+ \gamma \Psi _{i,j}^{(1)}+\zeta _{i,j}^{(2)}
+ \frac{{{m_1}}}{{{m_2}}}\frac{1}{{{\kappa_1 ^2}}}\left( {\frac{{{U_{x;\ i,j}}\Psi _{i + 1,j}^{(2)} + U_{x;\ i - 1,j}^*\Psi _{i - 1,j}^{(2)} - 2\Psi _{i,j}^{(2)}}}{{h^2}}}+ {\frac{{{U_{y;\ i,j}}\Psi _{i,j+1}^{(2)} + U_{y;\ i ,j-1}^*\Psi _{i ,j-1}^{(2)} - 2\Psi _{i,j}^{(2)}}}{{h^2}}} \right) ,
\end{array}
\end{equation}

\begin{equation}\label{eqs4n10}
\begin{array}{l}
{\partial _t}{U_{x;\ i,j}} =  - \frac{i}{\sigma }{U_{x;\ i,j}}{\mathop{\rm Im}\nolimits} \left\{ {\frac{{{W_{z;\ i,j}} - {W_{z;\ i,j - 1}}}}{{h^2}}+ {U_{x;\ i,j}}\Psi _{i,j}^{(1)*}\Psi _{_{i + 1,j}}^{(1)} + \frac{{{m_1}}}{{{m_2}}}{U_{x;\ i,j}}\Psi _{i,j}^{(2)*}\Psi _{_{i + 1,j}}^{(2)}}+\zeta _{i,j}^{(U_x)} \right\},
\end{array}
\end{equation}

\begin{equation}\label{eqs4n11}
\begin{array}{l}
{\partial _t}{U_{y;\ i,j}} =  - \frac{i}{\sigma }{U_{y;\ i,j}}{\mathop{\rm Im}\nolimits} \left\{{- \frac{{{W_{z;\ i,j}} - {W_{z;\ i,j - 1}}}}{{h^2}} + {U_{y;\ i,j}}\Psi _{i,j}^{(1)*}\Psi _{_{i,j+1}}^{(1)} + \frac{{{m_1}}}{{{m_2}}}{U_{y;\ i,j}}\Psi _{i,j}^{(2)*}\Psi _{_{i,j+1}}^{(2)}}+\zeta _{i,j}^{(U_y)} \right\}.
\end{array}
\end{equation}
\end{widetext}
Equations (\ref{eqs4n8}), (\ref{eqs4n9}), (\ref{eqs4n10}) and (\ref{eqs4n11}) are integrated by the forward Euler method.

Vortices are introduced through the boundary condition
\cite{Doria89}:
\begin{equation}\label{eqs3n8}
\mathbf{A}(r+L_l)=\mathbf{A}(r)+\nabla\chi_l, \ \ \ \  \Psi_k(r+L_l)=\Psi_k(r)\exp(i2\pi\chi_l/\Phi_0),
\end{equation}
with $l=x, y$ and $\chi_x=H_a L y$ and $\chi_y=0$. $H_a$ is the
applied magnetic field and should obey the vortex quantization
condition via $\oint d\mathbf{l}\cdot\mathbf{A}=2 m\Phi_0$ with an
integer $m$, which yields $H_a=2m\Phi_0/L^2$. Here $\Phi_0=hc/2e$ is
the flux quantum.

\subsection{Vortex structure}

\begin{figure}
\epsfysize=21cm \epsfclipoff \fboxsep=0pt
\setlength{\unitlength}{1cm}
\begin{picture}(6,21.0)(0,0)
\epsfxsize=7.6cm\epsfysize=6.0cm
\put(-0.5,15.0){\epsffile{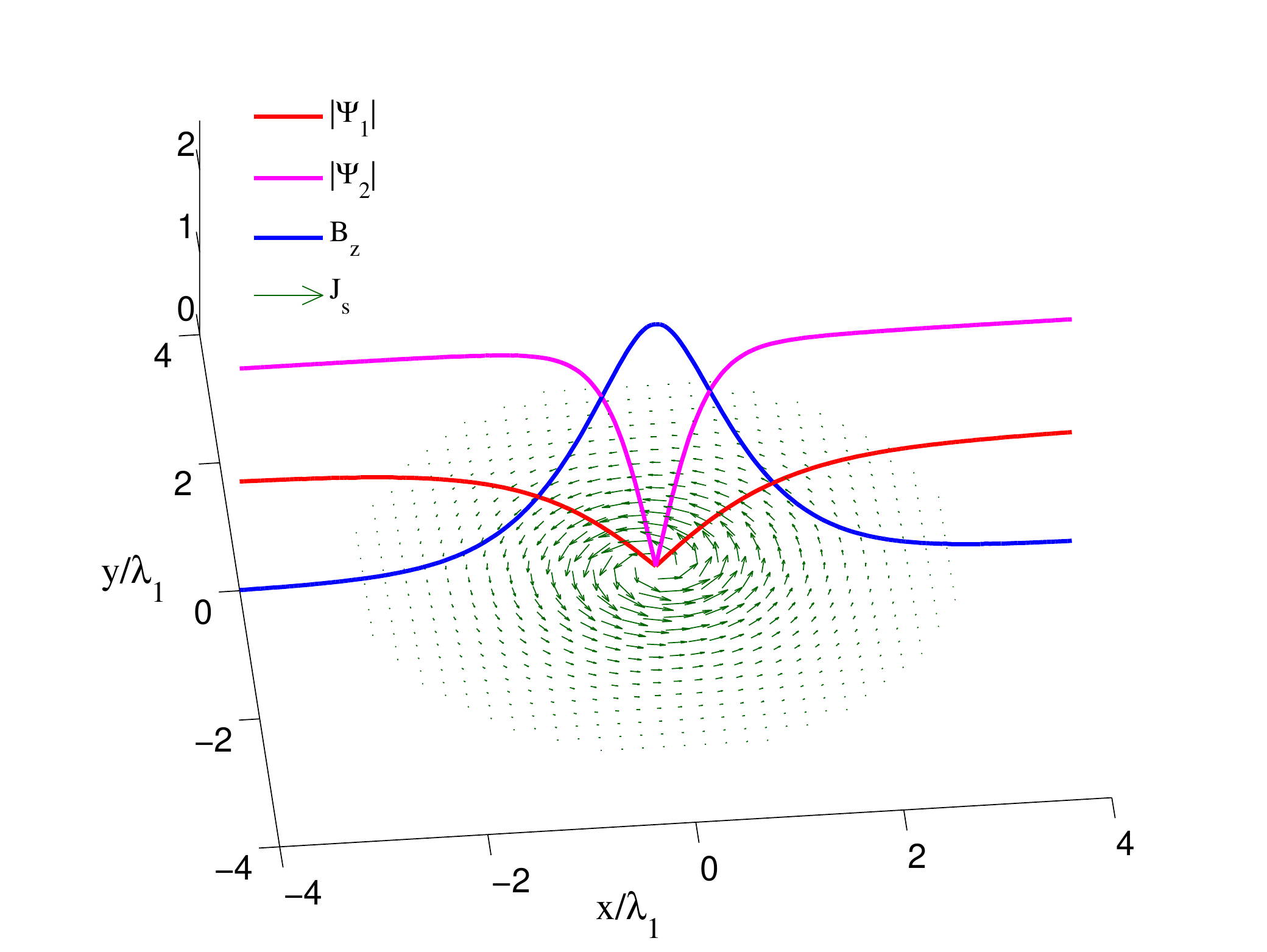}}
\epsfxsize=7.6cm\epsfysize=15.0cm
\put(-0.5,0.0){\epsffile{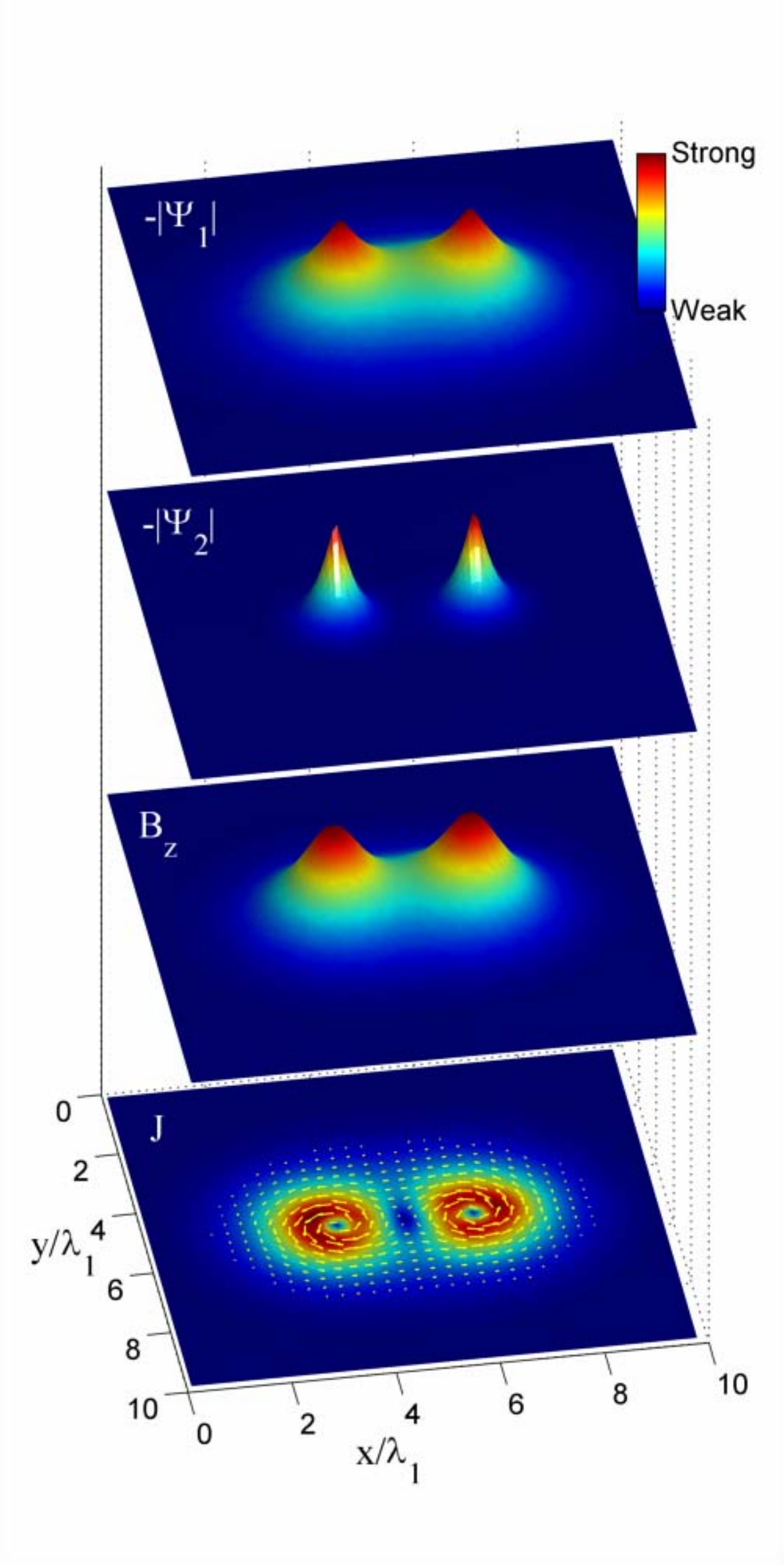}}
\put(0.0,20.8){(a)}
\put(0.0,13.8){(b)}
\end{picture}
\caption{\label{f0} (color online). (a) Profiles of order
parameters, magnetic field and supercurrent for a single vortex. (b)
A stable two-vortex solution in a finite sample derived numerically
using the approach of TDGL equations. The vortex separation is
$r_{\rm m}=2.7\lambda_1$.}
\end{figure}

We minimize $\mathcal{F}$ in Eq. (1) by solving the TDGL
equations numerically under the boundary conditions (\ref{eqs3n8})
specifying the number of vortices in the system. The structure of a
single vortex obtained by the TDGL equations for $m=1$ is shown in
Fig.~\ref{f0}(a). Three characteristic length scales are evident.

\subsection{Vortex attraction}

We then introduce two vortices into a square disk with size
$10\lambda_1\le L\le40\lambda_1$. The disk is large enough for one
to neglect the finite-size effect. We find a solution with two
vortices separated by $r_{\rm{m}}=2.7\rm{\lambda_1}$, independent on
$L$, as shown in Fig.~\ref{f0}(b). This indicates
unambiguously an attractive interaction at large separation and a
repulsive interaction at short distance between vortices. The
minimal energy associated with this vortex separation is therefore
demonstrated to be an intrinsic property of the superconductor. We
also confirm the stability of the above vortex solution against
thermal fluctuations in the present TDGL approach.

The size of the magnetic flux lies between the radii of the normal
cores associated with the two order parameters. As two vortices
approaching each other, they attract first by overlapping their
normal cores associated with $\Psi_1$ as shown in
Fig.~\ref{f0}(a). When the two vortices get closer, strong repulsion
caused by the magnetic flux sets in. An equilibrium configuration is
reached by compromising the attraction and repulsion, where the
normal core associated with $\Psi_i$ is overlapping strongly while
$\Psi_2$ is not as shown in Fig.~\ref{f0}(b). Thus vortices attract
at large separation and repel at short distance. It is the competition between the sizes of the normal cores
associated with the two components and the area of magnetic flux accompanying the vortex
cores that governs the interaction between vortices.

\section{Variational Calculations}

To obtain the full dependence of the interaction energy on the
vortex separation, one needs to fix vortices at desirable positions.
Since vortex is of structure and not a point object,
evaluation of vortex interaction at a given distance beyond the
London limit is not straightforward \cite{Nordborg00}. In
order to overcome this difficulty we resort to the variational
method developed by Jacobs and Rebbi \cite{Jacobs79} and generalize
it to two-band superconductors. The essence of this approach is to
construct "good" trial functions for two vortices at given
separation.

\subsection{Vortex structure}
\begin{figure}[t]
\psfig{figure=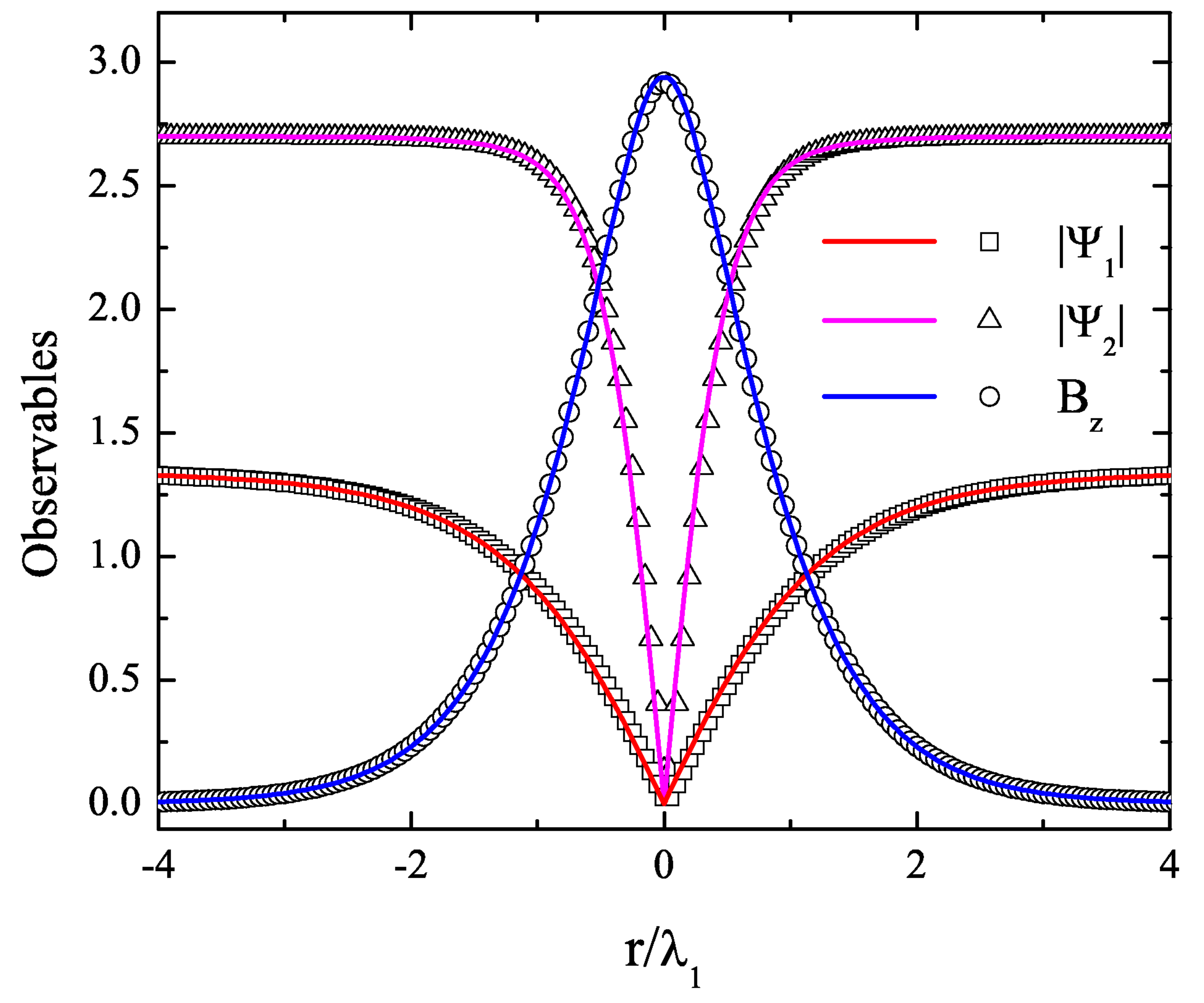,width=\columnwidth} \caption{\label{sf0}
(color online). Vortex structure obtained by the TDGL equations
(symbols) and the variational calculations (lines).}
\end{figure}
To reproduce the asymptotic behavior of a vortex at
$r\rightarrow\infty$, we use the following trial functions for one
vortex
\begin{equation}\label{eqs2n1}
f_1(r)=\sqrt{1+\gamma  \eta }+\exp \left(-\frac{r}{\sqrt{2}\xi_v}\right)\sum _{l=0}^n\left(f_{1,l}\left.r^l\right/l!\right),
\end{equation}
\begin{equation}\label{eqs2n2}
f_2(r)=\sqrt{\frac{\beta _1}{\beta _2}\left(\frac{\alpha _2}{\alpha _1}+\frac{\gamma }{\eta }\right)}+\exp \left(-\frac{r}{\sqrt{2}\xi_v}\right)\sum _{l=0}^n\left(f_{2,l}\left.r^l\right/l!\right),
\end{equation}
\begin{equation}\label{eqs2n3}
a(r)=1+\exp \left(-\frac{r}{\lambda_v}\right)\sum _{l=0}^n\left(a_l\left.r^l\right/l!\right) ,
\end{equation}
where $f_{1,l}$, $f_{2,l}$ and $a_l$ are variational parameters. In
the limit of $r\rightarrow0$, $f_i=0$ and $a\rightarrow r^2$
according to Eqs. (\ref{eqs1n9}), (\ref{eqs1n10}) and
(\ref{eqs1n11}). We have $f_{1,0}=-\sqrt{1+\gamma  \eta }$,
$f_{2,0}=-\sqrt{\frac{\beta _1}{\beta _2}\left(\frac{\alpha
_2}{\alpha _1}+\frac{\gamma }{\eta }\right)}$, $a_0=-1$ and
$a_1=a_0/\lambda_v$. For a giant vortex with vorticity equal to $2$,
we have $f_{1,1}=f_{1,0}/(\sqrt{2}\xi _v)$ and
$f_{2,1}=f_{2,0}/(\sqrt{2}\xi _v)$ in addition. Other coefficients are variational parameters to be determined numerically.

Denote the whole set of variation parameters $f_{1,l}$,
$f_{2,l}$, $a_l$ and so on by ${\bf V}$.
The GL free energy is of fourth order in $V_i$ and has the form of
\begin{equation}\label{eqs2n4}
\begin{array}{l}
\mathcal{F}=\mathcal{F}_0+\sum _i\mathcal{F}_i^{(1)}V_i+\sum _{i\geq j}\mathcal{F}_{{ij}}^{(2)}V_iV_j+\sum _{i\geq j\geq k}\mathcal{F}_{{ijk}}^{(3)}V_iV_jV_k\\
+\sum _{i\geq j\geq k\geq l}\mathcal{F}_{{ijkl}}^{(4)}V_iV_jV_kV_l.
\end{array}
\end{equation}
We use the Newton method of optimization\cite{NewtonOptimization} with the iteration procedure
\begin{equation}\label{eqs2n5}
V_i^{(m+1)}=V_i^{(m)}-\sum _j\left[\mathbf{H}^{-1}\right]{}_{ij}D_j^{(m)},
\end{equation}
where the superscript $(m)$ represents the value at the $m$th step.
$D_i=\partial \mathcal{F}\left/\partial _{V_i}\right.|_{V_i=V_i^{(m)}}$,
and the Hessian matrix $H_{ij}=\partial ^2\mathcal{F}/\partial V_i\partial V_j|_{V_{i,j}=V_{i,j}^{(m)}}$.
The stationary solution to Eq. (\ref{eqs2n5}) corresponds to the (local) minimum of the free energy.

Following the procedure of Eq. (\ref{eqs2n5}), we obtain the
variational coefficients, from which we can construct the vortex
solution. We truncate the higher-order corrections to the trial functions at $n=6$ and find the solution of a single vortex
with vorticity $1$. The results reproduce well those obtained by the
direct minimization of the GL free energy functional as shown in
Fig. \ref{sf0}.

\begin{figure*}[t]
\psfig{figure=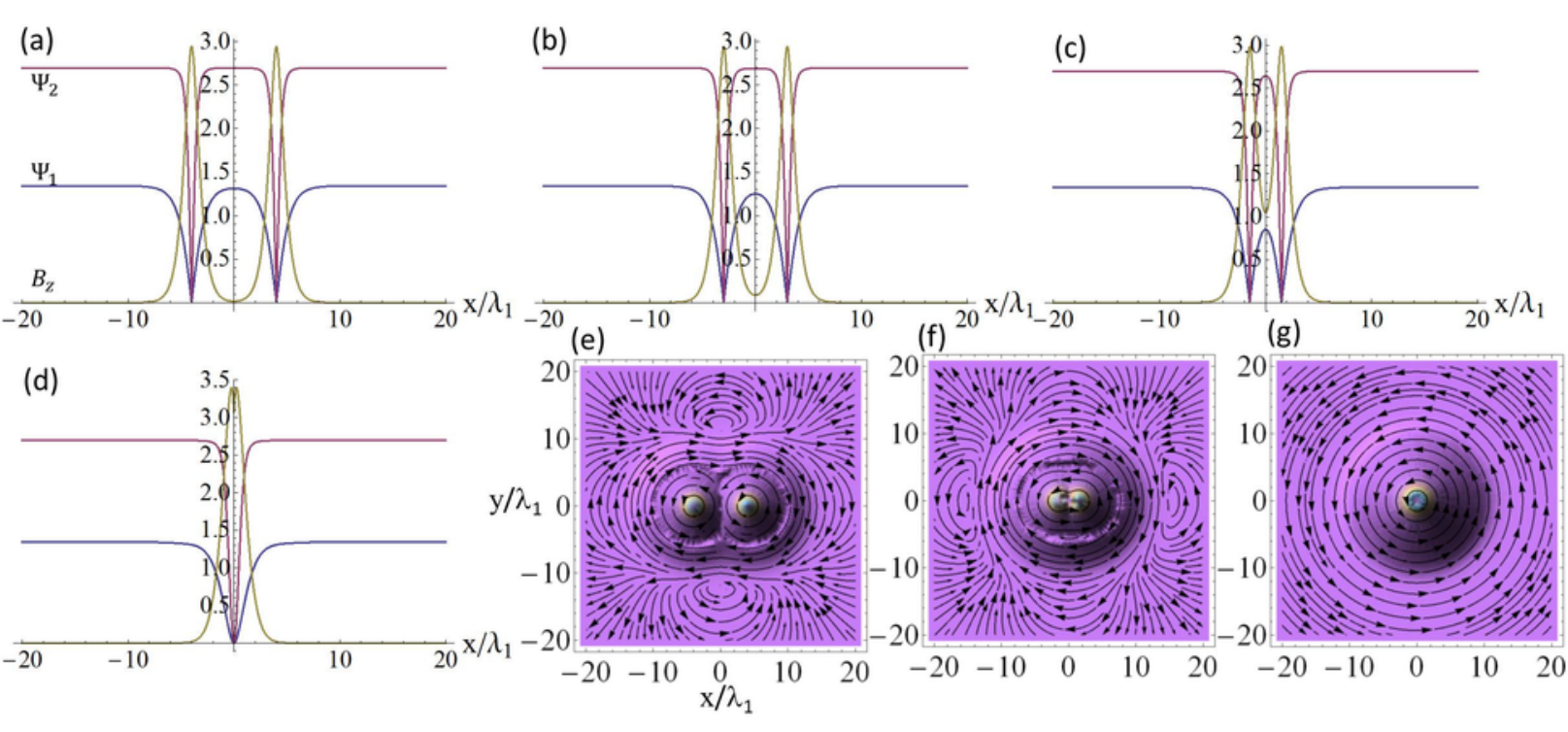,width=18cm} \caption{\label{sf1} (color online).
Distribution of the order parameters (purple and blue lines) and
 magnetic field (yellow line) at vortex separation  (a) $d=8\lambda_1$, (b) $d=6\lambda_1$,
 (c) $d=3\lambda_1$, and (d) $d=0\lambda_1$. Distribution of the supercurrent at vortex
 separation (e) $d=8\lambda_1$, (f) $d=3\lambda_1$, and (g) $d=0$.}
\end{figure*}

\subsection{Vortex interaction}
We proceed to consider the interaction between two vortices. For convenience,
we introduce complex representation of the 2D coordinates $z=x+i y$, and then
the winding phase part in Eq. (\ref{eqs1n8}) becomes
$\exp (i \theta )=\sqrt{z\left/z^*\right.}$. To construct trial functions
for two-vortex solution, we follow Ref. \cite{Jacobs79} and use the conformal
transformation of the complex plane
\begin{equation}\label{eqs3n1}
z=(z')^2-(d/2)^2.
\end{equation}
It is straightforward to see that the origin in $z$ plane have two
images in the $z'$ plane at $\pm d/2$, and that when $z'$ varies by
$2\pi$, $z$ varies by $4\pi$. This means that we may map a
one-vortex solution in $z'$ to a two-vortex solution in the $z$
plane. The phase factor of the two-vortex solution with
vortices cores at $z=\pm d/2$ can be constructed by this transform.
We seek the wave functions of the form
\begin{equation}\label{eqs3n2}
\Psi_i \left(z,z^*\right)=\left\{\left[z^2-\left(\frac{d}{2}\right)^2\right]/\left[{z^*}^2-\left(\frac{d}{2}\right)^2\right]\right\}^{1/2}f_i\left(z,z^*\right),
\end{equation}
with $i=1,2$. The trial function for $f_i$ can be constructed by the
following consideration. For $d\rightarrow \infty$, two vortices
behave independently, while at $d\sim 0$ two vortices merge and form
one giant vortex with vorticity $2$. In addition, we need also a
term to describe the interaction between two vortices. The trial
function therefore can be constructed
\begin{equation}\label{eqs3n3}
\begin{array}{l}
f_i\left(z,z^*\right)=\omega  f_i^{(1)}\left(\left|z-\frac{d}{2}\right|\right)f_i^{(1)}\left(\left|z+\frac{d}{2}\right|\right)\\
+(1-\omega )\frac{\left|z^2-\left(\frac{d}{2}\right)^2\right|}{\left|z^2\right|}f_i^{(2)}(|z|)+\delta f_i\left(z,z^*\right),
\end{array}
\end{equation}
where $f_i^{(1)}$ and $f_i^{(2)}$ are single-vortex solution with vorticity $1$ and $2$ respectively obtained by the variational calculations in the previous section, and $\delta f_i$ accounts for the interaction. $\omega$ interpolates two independent vortices solution and one giant vortex solution. The factor in the second term at the right-hand-side of Eq. (\ref{eqs3n3}) is to ensure that the wave function vanishes at the vortex cores $z=\pm d/2$. The interaction contribution may be constructed as follows
\begin{equation}\label{eqs3n4}
\begin{array}{l}
\delta f_i \left(z,z^*\right)=\\
\left|z^2-\left(\frac{d}{2}\right)^2\right|\frac{1}{\cosh  \left(\sqrt{2}\kappa_1  |z|\right)}\sum _{l=0}^n\sum _{j=0}^l f_{i,lj}\frac{\left|z|^{2l}\right.}{2}\left[\left(\frac{z}{z^*}\right)^j+\left(\frac{z^*}{z}\right)^j\right].
\end{array}
\end{equation}
The first factor is again to make sure that wave function vanishes
at the vortex cores, and the second factor accounts for the fact that
the interaction vanishes when $z\rightarrow\infty$.

Using the similar reasoning parallel to the construction of the trial
functions of $\Psi_i$, we can obtain the trial function for $\mathbf{A}$
\begin{equation}\label{eqs3n5}
\begin{array}{l}
\mathbf{A}=\omega \left[\frac{ i}{\kappa_1  \left(z^*-d/2\right)}a^{(1)}\left(\left|z-\frac{d}{2}\right|\right)+\frac{ i}{\kappa_1  \left(z^*+d/2\right)}a^{(1)}\left(\left|z+\frac{d}{2}\right|\right)\right]\\
+\frac{2 i}{\kappa_1  z^*}(1-\omega )a^{(2)}(|z|)+\delta a\left(z,z^*\right),
\end{array}
\end{equation}
where $a^{(1)}$ and $a^{(2)}$ are for the single-vortex solutions
with vorticity $1$ and $2$. The interaction contribution has the
form
\begin{equation}\label{eqs3n6}
\delta a\left(z,z^*\right)=\frac{1}{\cosh (|z|)}\left[z a_1\left(z,z^*\right)+z^*a_2\left(z,z^*\right)\right],
\end{equation}
with
\begin{equation}\label{eqs3n7}
a_k\left(z,z^*\right)=\sum _{i=0}^n\sum _{j=0}^ia_{{k,ij}}\frac{\left|z|^{2i}\right.}{2}\left[\left(\frac{z}{z^*}\right)^j+\left(\frac{z^*}{z}\right)^j\right],
\end{equation}
with $k=1, 2$.

Once the trial functions are constructed, we can do the optimization
according to Eq. (\ref{eqs2n5}). We put two vortices at $(-d/2, 0)$
and $(d/2, 0)$, and then calculate the distribution of the wave
functions, magnetic field and supercurrent. As displayed in
Fig. \ref{sf1}, when two vortices approach from $d=8\lambda_1$, the
condensate in the first band overlaps first, which causes attraction
between vortices, see Figs. \ref{sf1} (a) and (b). As $d$ is reduced
further, the magnetic fields start to overlap and strong
repulsion sets in as shown in Fig. \ref{sf1}(c). Finally, two
vortices coalesce into a giant vortex with vorticity $2$, see Fig.
\ref{sf1} (d).

\begin{figure}[b]
\psfig{figure=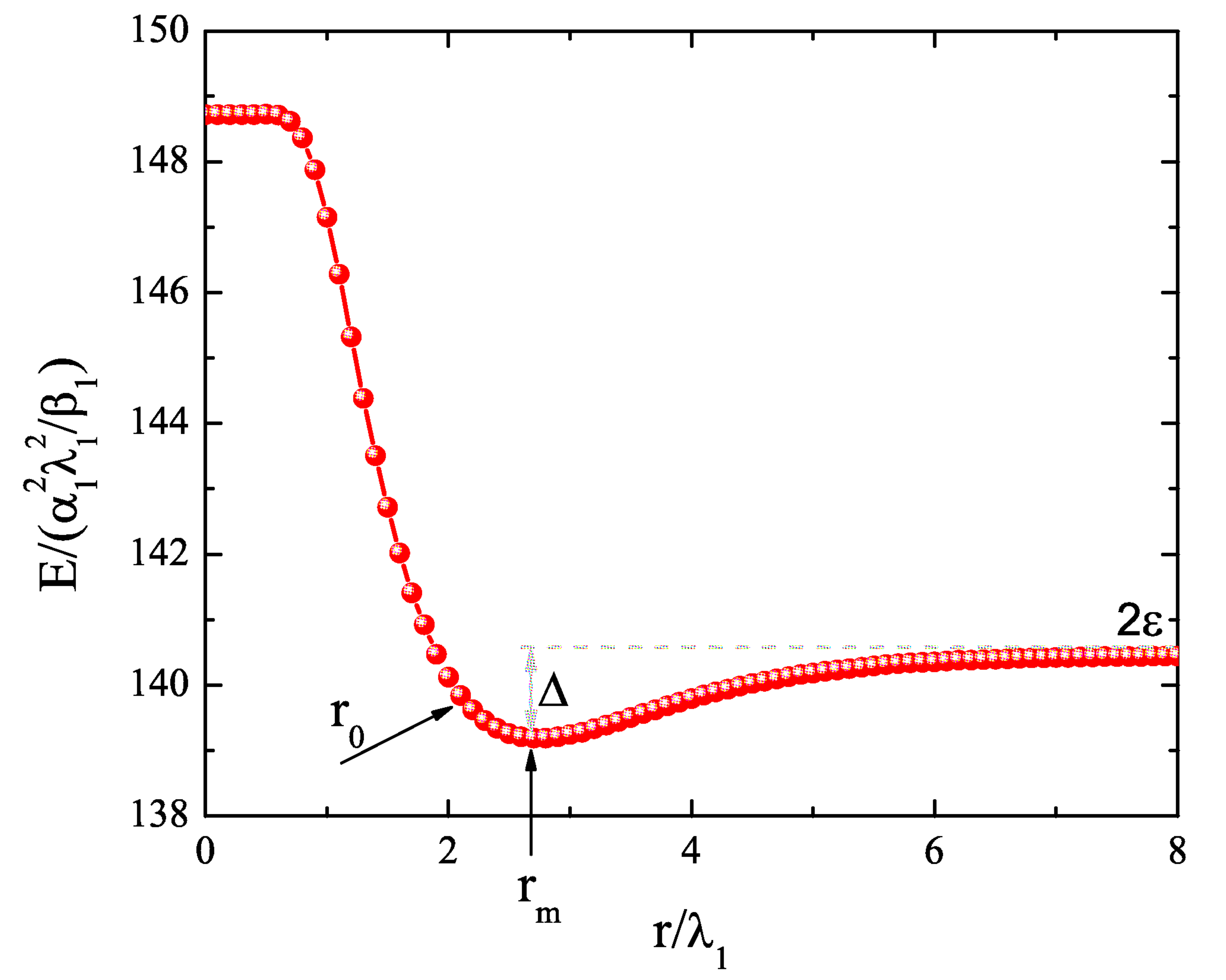,width=\columnwidth} \caption{\label{f1}
(color online). Dependence of the total energy $E$ per unit length
on the separation $r$ between two vortices. The distance $r_m$
corresponds to the energy minimum of the two-vortex system. Another
length denoted by $r_0$ is the lattice constant when the ordered
vortex patterns are formed as discussed later.}
\end{figure}

The resultant separation dependence of interaction between
two vortices is displayed in Fig.~\ref{f1}. The agreement between
the estimates on the position of minimal energy derived by the
variational technique and by the TDGL equations (Fig.~\ref{f0}(b))
serves a successful check of the validity of the variational
calculations. At large separation $r$, the attraction decreases
exponentially and the saturated energy corresponds to the
self energies of two isolated vortices. For $r<r_{\rm m}$, the
repulsion increases sharply and the energy at $r=0$ corresponds to a
giant vortex with vorticity $2$.

\subsection{Comparison on numerical approaches}
To calculate interaction between vortices poses a challenge to theory
since vortices are extended objects. In the London limit, the interaction
of vortices has been calculated analytically\cite{TinkhamBook}. For
general cases, one has to introduce constraints to fix two vortices at
a desired separation. A legitimate procedure is to fix vortices through
the boundary condition Eq. (\ref{eqs3n8}). Vortices in a disk of type II
superconductor tend to keep away from one another, but
they cannot leave the disk as imposed by the boundary condition. By
minimizing the free energy, we obtain the interaction energy and the
\emph{equilibrium} vortex separation for a given system size $L$. Gradually increasing
$L$, we obtain the interaction energy versus the vortices separation.
However, for vortices with attraction at large distance and repulsion
at short distance, this approach cannot give the dependence of the
interaction energy on vortex separation since the distance between two
vortices is always fixed, corresponding to the minimum of the interaction
energy. We notice that this approach is still the most legitimate way to prove
the existence of attraction at large distance and repulsion at short distance.

In numerical calculations of the GL free energy, one might
alternatively introduce pinning to vortices by fixing the amplitude and/or phase
of superconductivity order parameter in a certain region near the
vortex cores \cite{Misko03}. However, this method may introduce
artifacts when two vortices are close to each other since the the
order parameters change dramatically near the vortex
core. Furthermore, the local constraints are sometime insufficient to pin
vortices when the interaction becomes strong when vortices are close.
In contrast, in the variational approach shown above the global
structure (such as core of the vortices and asymptotic
behavior far from the core) of the vortices is maintained, and one
only adjusts the detailed structure through the variational
calculations. If the trial functions are appropriately chosen, the
variational approach gives superior results.

\section{Phase transition and phase diagram of vortex states}

\subsection{Vortex configuration}

The novel interaction profile shown in Fig.~\ref{f1} is expected
to dominate the vortex state of a macroscopic system, and
thus the phase diagram. In order to elucidate the situation, we
perform numerical simulations using the overdamped Langevin
dynamics \cite{AllenBook}
\begin{equation}\label{eqs5n1}
\eta d\textbf{r}/dt=-\partial E/\partial r+\textbf{F}^{(n)},
\end{equation}
where $E$ is the pair potential in Fig.~\ref{f1}, and
$\textbf{F}^{(n)}$ is the white noise force with $\langle F^n\rangle
=0$ and $\langle F_{i}^{(n)}(t)F_{j}^{(n)}(t')\rangle =2 T \eta
\delta_{i,j}\delta(t-t')$ with $T$ being the temperature and $\eta$
the viscosity.

Initially vortices are randomly distributed corresponding to
a high temperature and the system is annealed to $T=0$ by gradually
decreasing $T$, which yields the ground state. Equation
(\ref{eqs5n1}) is solved by the 2nd order Runge-Kutta method and the
simulation box is divided into cells with a cutoff radius $r_c$ to
accelerate the simulation. We fix the number of vortices $N_v$ and
set the simulation box with aspect ratio $2:\sqrt{3}$ to
accommodate the triangular lattice. The magnetic induction is given
by $B=2N_v\Phi_0/\sqrt{3}L^2$ with $L$ being the length of the
simulation box. Periodic boundary condition are used and $dt=0.02$,
$r_c=7.8$ and $N_v=400$. The results are checked
successfully by using finer $dt$, larger cutoff and more
vortices.

\begin{figure*}[t]
\psfig{figure=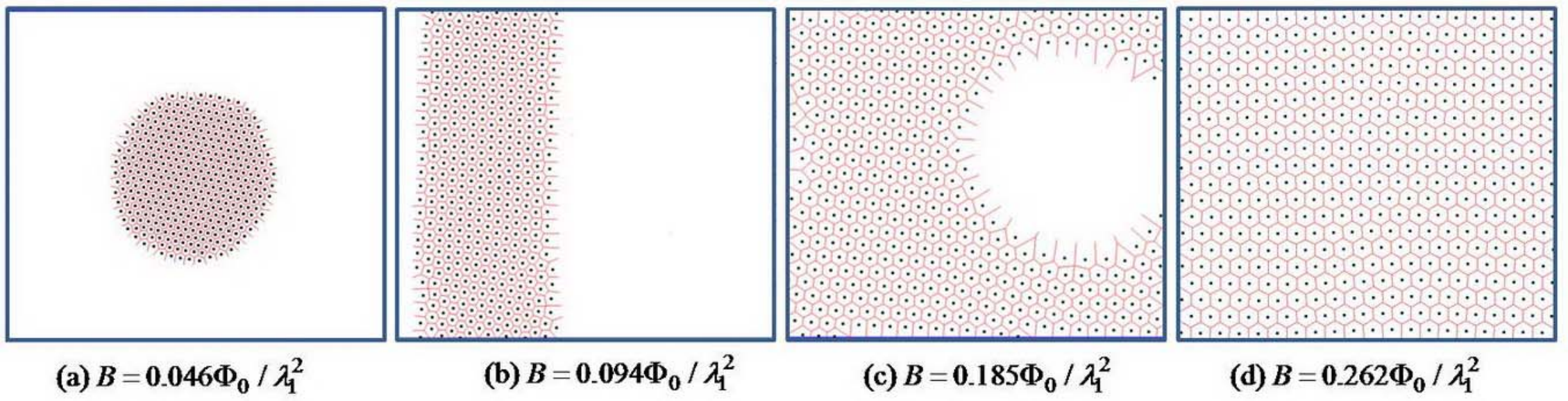,width=18cm} \caption{\label{sf4} (color
online). Vortex configurations at several typical values of magnetic
field: (a) vortex droplet, (b) vortex stripe, (c) vortex void, and
(d) triangular vortex lattice. Black dots denote the centers of
vortices and open region is the superconducting region. Red lines
are Voronoi polygons. The size of simulation box is (a)
$L=100\lambda_1$, (b) $L=70\lambda_1$, (c) $L=50\lambda_1$, and (d)
$L=42\lambda_1$, with the number of vortices $N_v=400$ fixed.}
\end{figure*}

A typical ground-state vortex configuration for small average
magnetic induction is presented in Fig.~\ref{sf4}(a). A circular
droplet of vortices with internal triangular order is obtained. This
vortex configuration is clearly determined by the vortex
interaction. In the presence of attraction, vortices prefer to stay
together to form a cluster.  For a vortex at the cluster surface,
the number of neighbors is smaller than that inside the cluster, and
thus it bares a higher energy. This results in a positive surface
tension for the vortex cluster, similar to type I superconductors.
The circular cluster minimizes the surface energy. On the other
hand, the repulsion force at short distance prevents the vortices
from collapsing. Due to the repulsive force, vortices inside the
cluster are triangularly ordered, same as type II superconductors.
The circular cluster of the closest packing triangular lattice
minimizes the total free energy. The lattice constant $r_0$ is
slightly smaller than $r_{\rm{m}}$ (see Fig.~\ref{f1}) due to the
contributions from vortices at large separations where interaction
is attractive.

Besides the circular droplet of vortices, vortex stripe and vortex void
are also observed at intermediate densities which minimize the free
energy for the system with given (finite) size and number of vortices. At a
small vortex density, the droplet phase [Fig. \ref{sf4}(a)] with
triangular order is stable. Upon increasing the magnetic field, the
vortex droplet expands, and at a threshold field a vortex stripe
phase [Fig. \ref{sf4}(b)] is preferred. The stripe then evolves into
vortex void configuration [Fig. \ref{sf4}(c)] when $B$ is increased
further. When $B$ becomes even larger, the void structure disappears
and a perfect triangular lattice [Fig. \ref{sf4}(d)] emerges and remain
stable until
the superconductivity is broken completely at $H_{c2}$.

Since the stable vortex configuration presumes the minimal surface
area, the transition fields between two configurations can be evaluated by comparing the
surface areas associated with the configurations. For a vortex droplet,
the surface energy is $2\pi R\sigma_s$ with the
radius of the cluster $R=\sqrt {\sqrt 3 N/2\pi}$ and the surface
tension $\sigma_s$. For a stripe configuration, the surface energy
is $2L\sigma_s$. The energy consideration gives the transition field
at $B_{\rm{ds}}={{{\Phi _0}}}/{{\pi r_0^2}}$. Similar argument gives
the fields of other structure transitions: transition from vortex stripe to
vortex void at $B_{\rm{sv}}={(2\pi \sqrt 3  -
3)\Phi_0}/{(3\pi r_0^2)}$, and transition from vortex void to vortex lattice at
$B_{\rm{vl}}={2{\Phi _0}}/{{\sqrt 3 r_0^2}}$. The numerical results
are consistent with these analytical estimates. It is noted that
these transition fields depend on the shape of the simulation box.
In the thermodynamic limit, the circular vortices droplet is the
only ground state at low magnetic field.

\subsection{Phase transition and phase diagram}

The phase separation between the Meissner region and a vortex
cluster indicates unambiguously a first-order phase transition, at
which clusters of vortices penetrate into the system upon increasing
the external magnetic field \cite{Babaev05}. The transition magnetic
field $H_{c1}$ is given by $H_{c1}=4\pi(\epsilon-n_1
\Delta)/\Phi_0$, with $n_1\simeq 3$, $\epsilon$ and $\Delta$ defined
in Fig.~\ref{f1} for the energy of a single vortex line and the
energy drop associated with the vortex attraction per unit length. The contribution from the
attraction component is small since $\epsilon\gg \Delta$ (see
Fig.\ref{f1}), and therefore $H_{c1}$ is close to that of a type II
superconductor with flux-line energy per unit length $\epsilon$. For
large magnetic fields a uniform vortex lattice of triangle
(Fig.\ref{sf4}(d)) has been observed same as type II superconductors.

\begin{figure}[b]
\psfig{figure=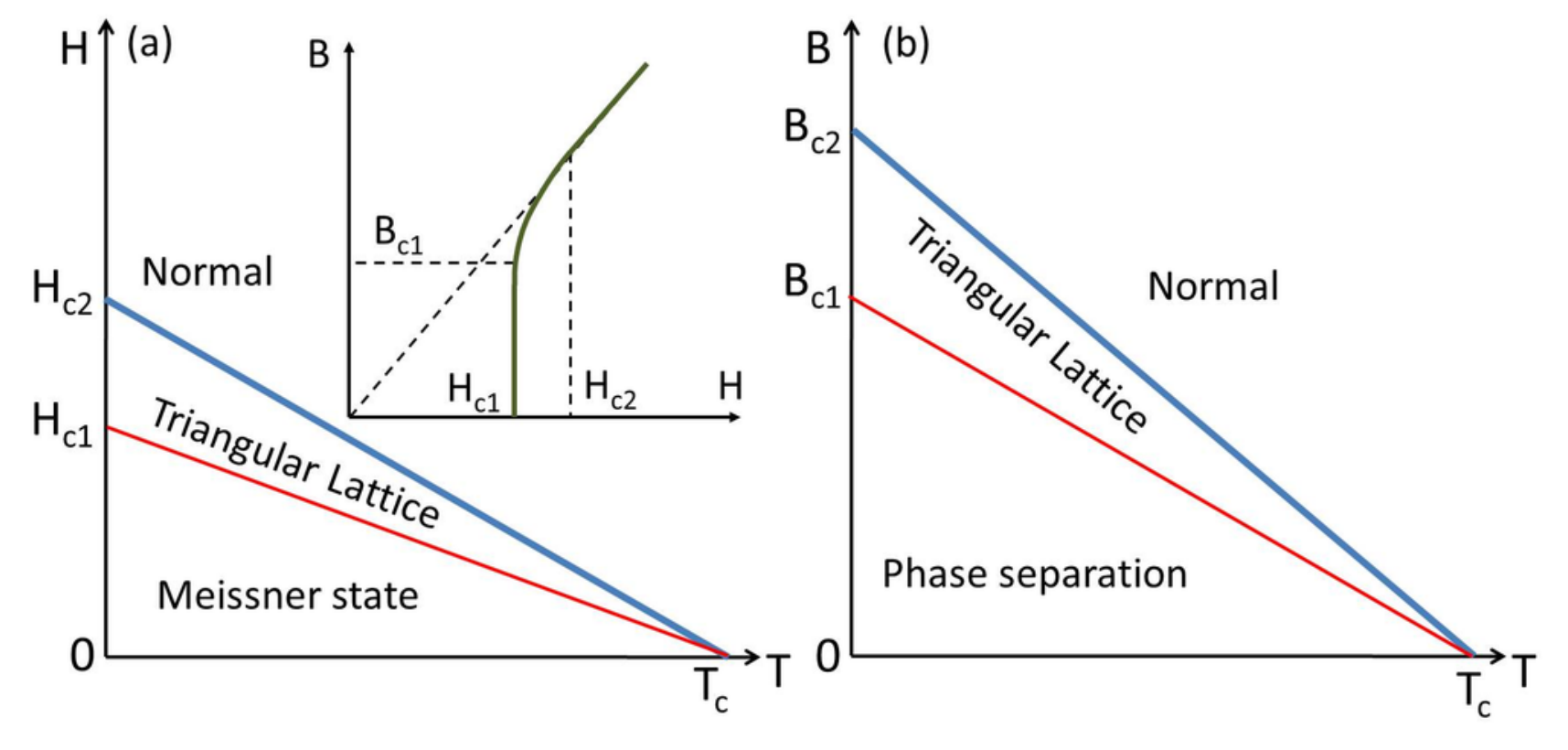,width=\columnwidth} \caption{\label{f3}
(color online). Mean-field $H-T$ (a) and $B-T$ (b) phase diagrams of
superconductors with competing inter-vortex interaction. The red
(thin) lines indicate the first-order phase transition, and the blue
(thick) lines are for the second-order phase transition. Inset is
for the dependence of magnetic induction on the applied field.}
\end{figure}

Now we can construct a mean-field phase diagram of vortex states in the
two-component superconductors. The $H-T$ phase diagram is depicted in
Fig.\ref{f3}(a), where the upper critical field $H_{c2}$ is the same
of a type II superconductor determined solely by the condensate with
the smaller coherence length. The $B-T$ phase diagram is
given in Fig.\ref{f3}(b), with the boundary between the phase
separation and the uniform triangular vortex lattice given by
$B_{c1}=n_2\Phi_0/r^2_{\rm m}$ with $n_2\simeq 2/\sqrt{3}$.

It is illuminating to compare the response to applied magnetic field in the
present system with those in conventional type I or type II superconductors. For type II superconductors, an
extremely dilute vortex lattice penetrates into a superconducting
sample at $H_{c1}$ and the magnetization curve is continuous at the
penetration. For type I superconductors, the magnetic field
penetrates into the sample and breaks the superconductivity at $H_c$
associated with a jump in the magnetization curve. For superconductors with competing repulsive and attractive inter-vortex interaction, clusters of vortex lattice with given lattice
constant $r_0$ penetrate into the sample associated with a discontinuous
jump in magnetic induction from zero to $B_{c1}$. It then increases
gradually with the external magnetic field until $H_{c2}$ at which
the superconductivity is destroyed (see inset of Fig.\ref{f3}). Therefore, the magnetic behavior
of these superconductors interpolates those of the type I and
type II superconductors.

\subsection{Effect of thermal fluctuations}

It has been revealed that thermal fluctuations should be weak in
$\rm{MgB_2}$\cite{Koshelev05}. In the present case, the competition
between long-range attraction and short-range repulsion may enhance
thermal fluctuations, and warrants additional treatment. Phase
diagrams of particles with a hard-core repulsion plus an attractive
tail have been investigated intensively. It is known that the
first-order gas-lattice phase boundary starts from $T=0$ and
$\rho=0$, and that all other transitions, such as gas-liquid,
liquid-lattice and possible lattice-lattice transformations, take
place at temperature $k_{\rm B}T/E_a \sim{\rm O}(1)$, where $E_a$ is
the strength of attractive potential (see for example
\cite{Camp03}). In the present flux-line system, the typical flux
segment relevant to distortion of vortex lattice is estimated as
$L_z\simeq \sqrt{\epsilon a^2/\Delta}$ with $a\simeq r_{\rm m}$,
which minimizes the energy associated with the tilt modulus and
vortex interaction (see for example \cite{Crabtree97}). This gives
an effective attraction potential $e_{\rm eff}=
L_z\Delta=\sqrt{\epsilon\Delta a^2}$. To estimate $\epsilon$ and
$\Delta$, we neglect the inter-band coupling ($\gamma=0$) for
simplicity. The attraction is caused by the overlap of the
condensate $\Psi_1$, and thus has an order of the energy of normal
core $\Delta\simeq {\left( {\frac{{{\Phi _0}}}{{8\pi }}}
\right)^2}\frac{1}{{\lambda _1^2}}$. The energy per unit length of a
single vortex is contributed from the normal cores of two
condensates and the magnetic energy, and is given by
$\epsilon\simeq{\left( {\frac{{{\Phi _0}}}{{8\pi }}}
\right)^2}\frac{1}{{{\lambda ^2}}} + {\left( {\frac{{{\Phi
_0}}}{{4\pi \lambda }}} \right)^2}\ln \left( {\frac{\lambda }{{{\xi
_2}}}} \right)$. Assembling these results and taking into account
the mean-field temperature dependence for the lengths, the condition
$k_{\rm B}T\simeq e_{\rm eff}$ for various phase transitions to
occur \cite{Camp03} spells as $(1-T/T_c)\simeq f G_{i}$,
where $f\sim100$ for a system with coherence length and penetration depth of
the same order, ${G_{i}} \equiv \frac{1}{2}{\left(
{\frac{{{k_B}{T_c}}}{{H_{1c}^2\xi _1^3(0)}}} \right)^2}$ is the
Ginzburg index for the first component with $\xi_1(0)$ the coherence length at zero
temperature and $H_{1c}\equiv\sqrt{4\pi\alpha_1^2/\beta_1}$. Since
$G_{i}\sim 10^{-6}$ as shown in Ref. \cite{Koshelev05}, the thermal
fluctuations are weak except the very small regime close to $T_c$,
i.e. $1-T/T_c\sim 10^{-4}$, where gas-liquid, liquid-lattice
and lattice-lattice transformations may be possible.

Effects of thermal fluctuations have also been investigated
by Langevin dynamics. In order to avoid possible artifacts due to
insufficient annealing, we heat the system from the ground state
with vortex configurations such as that shown in Fig.~\ref{sf4}(a).
At finite temperatures the perimeter of the cluster wiggles while
both the cluster itself and the inner triangular order remain
stable. Only at temperatures close $T_c$, single vortices are
evaporated from the cluster surface. These simulation results are
consistent with the above estimate on thermal fluctuations in the
present system. Through all the simulations, we cannot find any
random vortex patterns, such as gossamer- or stripe-like ones
\cite{Moshchalkov09}.

\section{Discussions}

\begin{figure}[b]
\psfig{figure=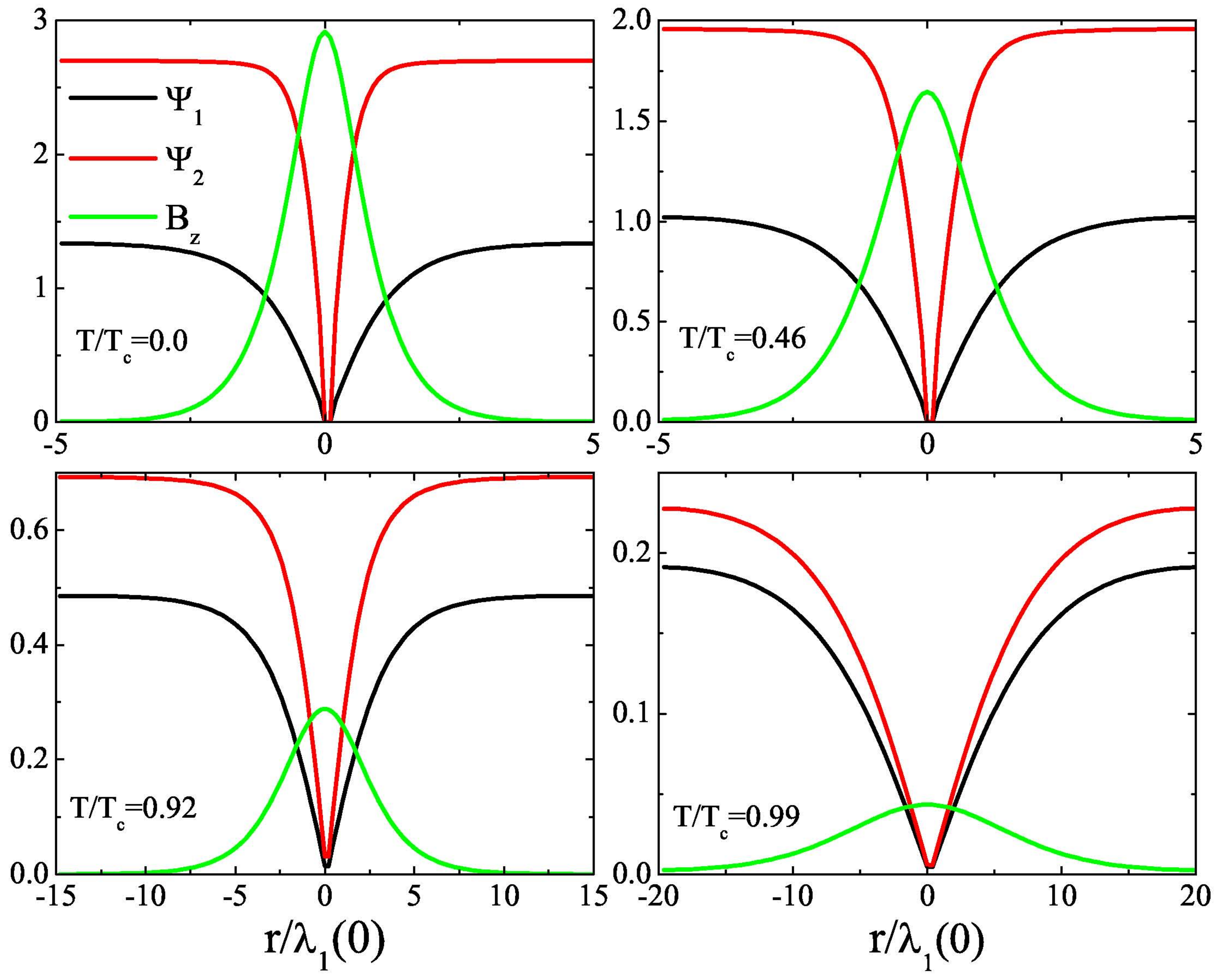,width=\columnwidth} \caption{\label{fT1}
(color online). Distribution of the order parameters and magnetic
field at several typical temperatures. The results are obtained with
$T_{c1}=0.6$, $T_{c2}=0.8$ and $T_c=0.87$.}
\end{figure}

Here we discuss a more general case that each condensate has
individual mean-field critical point $T_{ci}$ in absence of coupling
and a finite inter-band coupling. The inter-band scattering couples
two bands and enhances the critical temperature to $T_c>T_{ci}$. \cite{Kondo63}

For $T<T_{ci}$ and weak inter-band coupling $\gamma\ll1$, the
superconductivity is realized independently, and the physics is
qualitatively the same as that of two decoupled bands
$\gamma=0$.\cite{Babaev05} For $T_{c1}<T<T_{c2}$, the
superconductivity in the band $1$ is induced by the the band $2$
through the Josephson coupling. It is the regime discussed in Ref. \cite{Babaev10}.  When $T_{ci}<T<T_c$, the
superconductivity is purely induced by the inter-band coupling. The
physics can be quite different in different temperature regimes. We
notice that in principle the GL theory is valid only close to $T_c$,
and thus in the highest temperature region.

From Eq. (\ref{eqs1n1}), $T_c$ is given by the condition
$\alpha_1(T) \alpha_2(T)-\gamma^2=0$, where $\alpha_i(T)$
depends on the temperature and can be derived from the BCS theory
\cite{Gurevich07}. Close to $T_c$,
$\tau\equiv\frac{\gamma^2-\alpha_1 \alpha_2}{\alpha_1
\alpha_2}=(T-T_c)/T_c\ll1$, the order parameters up to the leading
order can be derived from Eqs. (\ref{eqs1n4}) and (\ref{eqs1n5}) for
a uniform superconductor \cite{Geyer10}
\begin{equation}\label{EqT1}
\Psi_{10}^2=\frac{\alpha_2^2\alpha_1\tau}{\alpha_2^2\beta_1+\alpha_1^2\beta_2},
\end{equation}
\begin{equation}\label{EqT2}
\Psi_{20}^2=\frac{\alpha_1^2\alpha_2\tau}{\alpha_2^2\beta_1+\alpha_1^2\beta_2}.
\end{equation}
We then consider some deviations from the bulk values $\Psi_{i}=\Psi_{i0}+\phi_{i}$.
From Eqs. (\ref{eqs1n4}) and (\ref{eqs1n5}), the deviations are governed by
\begin{equation}\label{EqT3}
\alpha _1\phi_1+3\beta _1\Psi _{10}^2\phi_1-\frac{\hbar^2}{2m_1} \nabla^2\phi_1-\gamma  \phi _2=0,
\end{equation}
\begin{equation}\label{EqT4}
\alpha _2\phi_2+3\beta _2\Psi _{20}^2\phi_2-\frac{\hbar^2}{2m_2} \nabla^2\phi_2-\gamma  \phi _1=0.
\end{equation}
Taking the solution of form $\phi_i\sim \exp[-x/(\sqrt{2}\xi_v)]$, we can derive the
coherence length which is divergent at $\tau \rightarrow 0$,
 $\xi_v=\sqrt{\left(\frac{\alpha _2}{4m_1}+\frac{\alpha _1}{4m_2}\right)\hbar ^2}\frac{1}{\sqrt{2\tau}}$. We emphasize that
there is only one coherence length for the two order parameters when
$\tau \rightarrow 0$. The London penetration depth is still given by
Eq. (\ref{eqs1n7}) with the order parameters given in Eqs.
(\ref{EqT1}) and (\ref{EqT2}). Calculations of the order parameters
near $H_{c2}$ and of the spatial correlation of
$\langle\Psi_i(\mathbf{r})\Psi_j(0)\rangle$ when thermal
fluctuations are present yield only one divergent coherence length
consistently. Therefore, the superconductors are either type I or type II.

Since it was discussed that vortex attraction originates from the
overlapping of normal vortex cores,\cite{Babaev05} let us check the temperature dependence of the structure
of a single vortex ranging from 0 to $T_c$. For simplicity, we take
$\alpha_i(T)=\alpha_i(0)(1-T/T_{ci})$ and assume $\beta_i$ and
$\gamma$ are $T$-independent. The parameters at $T=0$ are the same
as those in the previous sections. The vortex structure is shown in
Fig. \ref{fT1} for several typical temperatures. It becomes clear
that the sizes of vortex cores for $\Psi_1$ and $\Psi_2$ get closer
to each other as $\tau$ approaches 0.
Analytical calculations on the structure of the nonlinear
vortex core also reveal an identical size for the vortex cores
associated with the two components close to $T_c$.\cite{Hu11a} Therefore, the
interaction between vortices is purely repulsive in the present case.

This result is understandable since the inter-band coupling is dominant at $\tau\ll1$, and thus the superconductivity in the
two bands is strongly correlated and described by a unique coherence length. Nevertheless, if the inter-band coupling $\gamma(T)$ and $\alpha_i(T)$ all vanish at a single temperature $T_c$ as discussed in the previous sections, there will be two divergent
lengths and thus different core sizes for the two components. This permits vortices to exhibit the peculiar interaction profile shown in Fig. \ref{f1}. Therefore, we find that the non-monotonic inter-vortex interaction, and thus the peculiar phenomena
discussed above, only appear when the inter-band coupling and $\alpha_i(T)$ all vanish at $T_c$ within the framework of the GL theory.

\section{Conclusion}
We have investigated the interaction between vortices in two-band
superconductors. When the size of the magnetic flux of a vortex lies
between the sizes of the two normal cores associated with the two
condensates, the interaction between vortices is attractive at large
separation and repulsive at small separation. This was demonstrated
clearly by the time-dependent Ginzburg-Landau calculations on two
vortices in a superconducting disk. The equilibrium
distance between two vortices is independent of the size of the
disk for simulations, which clearly shows the existence of energy minimum in the
interaction potential. The full dependence of interaction energy on
the vortex separation is obtained by the variational calculations.
The two methods give the same estimate on the vortex separation with minimal free energy.

We have studied stable vortex configurations for a large number of vortices by
Langevin dynamics adopting the novel distance dependence of vortex
interaction. Circular vortex clusters coexisting with
Meissner phase are observed for small and intermediate vortex densities. The transition from
Meissner state into vortex states is therefore of first order associated with
a sharp increase of magnetization. The superconductivity associated with
the triangular vortex lattice is suppressed by a strong magnetic field
in the same way of a type II superconductor. Therefore, the magnetic behavior
of these superconductors as summarized in the mean-field phase
diagrams interpolates those of the type I and type II superconductors. In
most temperature regions except very close to the critical point, thermal
fluctuations are weak for small Ginzburg number, the same as single-band
superconductors.

The above interesting phenomena take place provided, first,
the bare mean-field critical points of the two bands
coincide with each other and the inter-band coupling vanishes at the same
temperature, which permit two divergent core sizes associated with the
two condensates even close to the critical point; second, the parameters
in the Ginzburg-Landau free energy are appropriate to achieve the
relations among the two coherence lengths and the penetration depth.

When two condensates have different bare transition temperatures and
the inter-band coupling is finite, the superconductivity is induced
by the inter-band scattering when temperature is sufficiently close
to the critical point of the composite system. In this case, superconductivity in
the two bands couples strongly, and only one divergent coherence length exists,
and thus the superconductors are either type I or type II.

\vspace{5mm}

\section{Acknowledgement }
The authors thank V.~Vinokur, A.~Gurevich, H.~Brandt, M.~Tachiki, and Z.~Wang
for discussions.  This work was supported by WPI Initiative on Materials
Nanoarchitectonics, and Grants-in-Aid for Scientific Research (No.22540377),
MEXT, Japan, and partially by CREST, JST.


\end{document}